\documentclass[preprint]{aastex}

\newcommand{\bc}{\begin{center}}
\newcommand{\ec}{\end{center}}
\newcommand{\be}{\begin{enumerate}}
\newcommand{\ee}{\end{enumerate}}
\newcommand{\beq}{\begin{equation}}
\newcommand{\eeq}{\end{equation}}
\newcommand{\beqa}{\begin{eqnarray*}}
\newcommand{\eeqa}{\end{eqnarray*}}
\newcommand{\bi}{\begin{itemize}}
\newcommand{\ei}{\end{itemize}}
\newcommand{\bt}{\begin{tabular}}
\newcommand{\et}{\end{tabular}}
\newcommand{\etal}{{et~al.\/}\ }
\newcommand{\eg}{{\em e.g.\/}\ }
\newcommand{\ie}{{\em i.e.\/}\ }

\newcommand{\ha}{$\rm H{\alpha} $ }

\newcommand{\mo}{ \cal  M_{\odot}  }
\newcommand{\kms}{$\rm km \ s^{-1}$}
\newcommand{\dghr} {30-Dor GHR's}

\newcommand{\ho}{$ \rm h^{-1} $ }
\newcommand{\y}{YPC's }
\newcommand{\g}{GC's }
\newcommand{\s}{GCS }
\newcommand{\sn}{\(\rm S_N \) }

\begin{document}
 
\slugcomment{Paper I: To be submitted to A.J.}

\shorttitle{NGC~3256: H~II Regions}
\shortauthors{English \& Freeman}

\title{Giant H~II regions in the merging system NGC~3256:  Are
they the birthplaces of globular clusters? }

\author{J. English}
\affil{University of Manitoba}
\authoraddr{Department of Physics and Astronomy, University of
Manitoba, Winnipeg, Manitoba, Canada R3T 2M8}

\author{K.C. Freeman}
\affil{Research School of Astronomy and Astrophysics, The Australian
National University}
\authoraddr{Mount Stromlo Observatory, Private Bag, Weston Creek P.O.,
  A. C. T. ~2611, Australia}

\begin{abstract}
CCD images and spectra  of ionized hydrogen in the merging system
NGC3256 were acquired as part of a kinematic study to
investigate the formation of globular clusters (GC) during the interactions
and mergers of disk galaxies.  This paper  focuses on the proposition
by \citet{kc88}  that giant H~II regions, with
an \ha luminosity $> \rm 1.5 \times 10^{40}\  erg \ s^{-1}$, are
birthplaces of young populous clusters (\y).

Although NGC~3256 has relatively few (7) giant H~II complexes, compared
to some other interacting systems, these regions are comparable in
total flux to about 85 30-Doradus-like H~II regions (\dghr).  The
bluest, massive \y \citep{zaeks95} are located in the vicinity of
observed \dghr, contributing to the notion that some fraction of \dghr
\ do cradle massive YPC's, as 30 Dor harbors R136.  If interactions
induce the formation of \dghr, the observed luminosities indicate that
almost 900 \dghr \ would form in NGC~3256 throughout its merger
epoch. In order for \dghr \ to be considered GC progenitors, this
number must be consistent with the specific frequencies of globular
clusters estimated for elliptical galaxies formed via mergers of
spirals \citep{az93}. This only requires that about 10\% of NGC~3256's
900 \dghr harbor \y which survive several Gyr and have masses $\geq \
\cal M_{\rm R136}$.

\end{abstract}
\keywords{galaxies: individual (NGC 3256) ---  galaxies: interactions ---
globular clusters: general}
\section{Introduction}
\label{intro}
\subsection{Motivation for Kinematic Observations}
\label{complement}

Based on observations of starburst activity and the large amounts of
molecular gas in interacting galaxies, \nocite{sch88} Schweizer (e.g.
1988) suggested that merger events provide an ideal environment for
the formation of massive star clusters. Theoretical support that these
would evolve into globular clusters (GC's) was provided by
\nocite{elmef97}Elmegreen \& Efremov~(1997) who argued that massive
{\em bound} clusters preferentially form in a high pressure
environment such as that created during a starburst.  Starbursts
frequently are observed in galaxy-galaxy interactions and mergers,
presumably enhanced by tidally induced shocks~\citep{lar87}.  Other
suggestions of how interactions may provide the dynamical phenomenon
necessary to produce young populous clusters (\y; also known as super
star clusters) and how these will subsequently evolve into \g have
been described, for example, by \nocite{elm93} Elmegreen (1993),
\nocite{bcm93}Boily, Clarke \& Murray (1993), and \nocite{ml93} Murray
\& Lin (1993).

HST searches of interacting and merging systems have been
spectacularly successful in finding compact blue objects which are
likely to be \y formed during the interactions and mergers.  Large
numbers of blue compact objects were detected in the interacting
galaxy NGC~4038/39~\citep{ws95}, and merger remnants
NGC~1275~(\citet{carl98}, \citet{fab93}, \citet{hfs92}) and NGC~7252
\citep{wsl93}. The brightest of  these
clusters have the attributes expected of proto-globular clusters
\citep{whitmore01}.

Underlying all of this is the hope that, as the merging systems settle
to form an elliptical galaxy, the \y will survive to contribute to
its rich globular cluster system (\s). An important observational
element is missing from this picture: are the kinematics of the \y
at the present time consistent with their ultimate spatial distribution in the
cluster system of the final elliptical remnant?  To answer this question, we
need to compare velocity data for the \y (or for their progenitors)
with the kinematics of associated features in the hydrodynamical
simulations of merging spirals.

To make this comparison, GC progenitors need to be
identified. Assuming \y evolve into GC then some possible sites for
GC formation during mergers include giant H~II regions and the
detached H~I fragments and gas clumps within the extended tidal
features found around merging galaxies (e.g.  \citet{dm94}).  We would
like to eventually test whether the spatial and velocity distributions
of these GC progenitor {\em candidates} are consistent with the
statistical distribution and kinematics of small-scale condensations
in the hydrodynamical models of merging galaxies.  We have collected data
relevant to measuring the masses and the spatial and velocity
distributions of H~I and H~II regions within galaxy systems that
sample the range of galaxy-galaxy interaction stages leading to a
merger remnant.  This paper, and a companion paper by \nocite{enfa95}
English et al.\/~(2002: Paper~II), describe observations of one
merging system, NGC~3256.

\subsection{Globular Cluster Formation Scenarios}
\nocite{sz78}Searle \& Zinn~(1978) proposed that \g form in small
fragments which are subsequently accreted by the parent galaxy. Their
fragments are presumably the low-mass end of the hierarchy of
fluctuations that aggregate to form a spiral system. In major mergers,
another source of fragments are those torn from  interacting galaxies;
falling back into the potential well of the remnant galaxy, they
experience star-formation due to tidally induced shocks, and contribute
their resultant \g to the emerging elliptical.  This globular cluster
formation scenario motivated the neutral hydrogen observations of NGC 3256
 which will be presented in Paper II.

The ionised hydrogen observations presented here were motivated by
\nocite{kc88}Kennicutt and Chu's (1988) suggestion that giant H~II
regions are the birthplaces of \y like the young globular clusters
seen in the Magellanic Clouds.  We consider the possibility that giant
H~II region formation is enhanced in the disks of interacting spirals
during the interaction.  The associated \y would be subsequently
redistributed, along with the other components of the two disks
involved, as the merger of the parent galaxies proceeds (e.g.
\citet{sch95}).  This is supported by the clustering of \y in
NGC~4038/39. The youngest star clusters spatially cluster together more than
the intermediate age clusters (which are apparently associated with tidal
tails) and than the old \g (distributed mostly throughout the disk),
suggesting mixing has occurred over several orbits~\citep{zfw2001}.
This is also consistent with the observation that the
clusters in giant ellipticals have a roughly similar spatial
distribution to the diffuse light distribution.  [Although in some
ellipticals the GC distribution is a little flatter than
the light distribution~\citep{har93}, in others like NGC~1399 the
cluster and light distributions are almost identical in
slope~\citep{hh87}.]

The mass spectrum of \g in ellipticals results from the initial mass
spectrum during the cluster formation epoch and the effects of
subsequent dynamical evolution.  Hence the observation that, in
elliptical galaxies, the mass spectrum of \g with masses greater than
$\rm 10^5 \mo$ is fit by a near universal power law \citep{harpud94}
which has the same slope as the giant H~II region mass spectrum
(e.g. \citet{keh89}; \citet{msu96}) is interesting in the context of
this globular cluster formation scenario.

\subsection{The Specific Frequency of Merger Remnants}
\label{specfreq}

The specific frequency \sn of globular clusters is the number of \g
per unit absolute magnitude $M_V = -15$ of galaxy luminosity
\citep{hvdb81}.  The range of \sn observed for elliptical galaxies
covers an order of magnitude \citep{har93} and its value for spiral
galaxies can be an order of magnitude lower than for cD ellipticals
which are at the high extreme.

\nocite{sch88}Schweizer (1988) proposed that the \sn of a remnant
elliptical would be larger than the value derived from the sum of the
\g populations of the parent galaxies due to \g formation during
mergers.  \nocite{az93}Ashman and Zepf~(1993) determined the number of
\g per unit stellar mass, using a characteristic ${\rm M/L_V}$ for
each morphological type.  They find that normal (\ie\ non-cD)
elliptical galaxies have more than twice as many \g per unit stellar
mass as spirals.  Therefore even the construction of normal
ellipticals via the merger of spirals would require the formation of
\g in the merger process.

Of course this would also require \g to form in preference to
non-cluster stars in the merger.  While this is theoretically
plausible (e.g. \citet{elmef97}, and vast numbers of \y are clearly
formed in mergers like NGC 4038/9, there is no direct evidence that
this is so. Still NGC 1275 provides an example of a system with a high
\sn ($\sim$27) and \nocite{carl98}Carlson \etal (1998) find that it
will remain larger than that of the old \g population even if the
majority of the low-mass clusters are destroyed.

We should comment here on an important paper by \nocite{forb97}Forbes
\etal\ (1997) who examined the color distribution of \g in
ellipticals. In galaxies with a high specific frequency of \g (mostly
cD ellipticals), they found that the excess \g are mostly
metal-poor. This is not consistent with forming the high specific
frequency of \g through mergers of spiral galaxies with near-solar
metallicity. But it does not argue against formation of \y in mergers:
this is clearly observed to happen.  The Forbes \etal\ result suggest
that the bulk of GC formation in cD ellipticals may have occurred
during the early phase of hierarchical merging, while the mean
metallicity was still low. That is, the \g may have formed
through the early merging of a large number of relatively unevolved
fragments, rather than through the later merging of more chemically
evolved galaxies.

\subsection{NGC~3256}

NGC~3256 is a nearby (2820 \kms) merging system which is experiencing
a spatially extended and highly luminous starburst ($\rm \sim 3 \times
10^{11}\ L_{\odot}$; \citet{ssp89}).  The $\rm 10 \ \mu m $ luminosity
of the visible \ha nucleus ($\rm \sim 2 \times 10^{10} \ L_{\odot}$)
rivals that of Seyfert galaxies~\citep{grah84}.  The system has
another ``nucleus'' 5 arcsec to the south, obscured by dust but
detected in non-optical wavelengths such as: 3 cm \citep{nf95};
near-infrared K-band (e.g. P. MacGregor (see Paper II), \citep{lip2000}); 
and X-ray \citep{lira02}.  This system also
has a star-forming complex in its `disk', an arc of giant H~II
regions, and 2 extended tidal tails.  The 2 X-ray nuclei, along with
the diffuse X-ray emission, support the scenario that NGC~3256 is
powered by a starburst rather than an AGN \citep{lira02}. Therefore we
tentatively assume that NGC~3256 has two cores (one from each parent
galaxy) that are on the verge of merging.

\nocite{lip2000} L\'ipari \etal (2002) argue that NGC~3256 is the
result of a multiple merger with the disk star forming complex as the
third nucleus.  Since there are only 2 tidal tails, they suggest that
either an on-going merger system encountered a third galaxy or that 
2 gas rich spirals plus a minor galaxy merged simultaneously.   These
cases, for our order of magnitude estimates, can be replaced by
a 2 galaxy prograde interaction scenario.

In Paper II we estimate the dynamical timescale associated with the
neutral hydrogen component of the tails and compare our observations
with numerical simulations of starbursts in galaxy mergers which
generate elliptical remnants (e.g. \citet{mh94}).  In these models the
timescale since the last pericentre approach of the parent galaxies
through to coalescence is about 500 Myr, comparable to our estimate of
the dynamical time ($\sim $ 500 \ho\ Myr) since pericentre in the
observed NGC~3256 system.  If the model parent galaxies contain dense
bulges, a strong starburst is also produced at about 500 Myr
post-pericentre.  Although the K-band luminosity profile of NGC~3256
is not yet the $r^{1/4}$ distribution of a fully relaxed system, the
profile is consistent with a merger phase in which the cores of two
galaxies are about to coalesce~\citep{mo94}.  Subsequent violent
relaxation would cause this system to have the structure of an
elliptical galaxy in its final merger state~\citep{sch86}.

\nocite{hb93} Hernquist and Bolte~(1993) examine the GCS's produced in
simulations of mergers of two disk galaxies.  The surface brightness
profile of the model GCS's are  also well fit by the observed r$^{1/4}$
law. They argue that one of the sites of new cluster formation will be
in the inner regions of the merger remnant, because a large fraction of
disk gas is driven to each disk's centre and subsequently the cores of
the disks merge~(\citet{bh92}, \citet{mh94}).  Since NGC~3256 is
approaching the merger stage (see the merger sequence
by~\citet{toom77}), it is thus not surprising that H~II region formation
is enhanced in the galaxy's disk.   Again,  the inner disk gas would
provide the high pressure environment to produce bound clusters
according to \nocite{elmef97} Elmegreen \& Efremov~(1997). 
And, though all may not be bound, \nocite{zaeks95} Zepf \etal (1999) 
find several hundred compact, blue clusters likely to be young
clusters within the 7 square arcsec 
central region.

Our photometry (\S~\ref{dat}) shows that, although NGC~3256 has
relatively few H~II regions compared to some other interacting
systems, these regions are comparable in flux to about 85
30-Doradus-like H~II regions (\S~\ref{results}).  Since the 30 Doradus
nebula contains the YPC NGC~2070~(\cite{mey93}), we label as ``\dghr
'' any giant H~II regions with luminosities greater than or equal to
$\rm 1.5 \times 10^{40}\ erg \ sec^{-1}$~\citep{kc88}.  We present
2-D spectra (\S~\ref{dat}) used to compare the velocity distribution
of H~II regions in this merging system with numerical models; these
extend spatially beyond the spectra of \citep{lip2000}
(\S~\ref{results}).  We estimate the number of \y which could be born
between the last closest approach of the parent galaxies and the
currently observed epoch in \S~\ref{results}.  We summarize in
\S~\ref{concl} that NGC~3256 is an emerging elliptical galaxy that is
producing a population of \y consistent with the specific frequencies
of \g that would be expected for elliptical remnants formed via
mergers of spiral galaxies~\citep{az93}.


\section{Observations, Reductions, and Data Analysis}
\label{dat}
\subsection{CCD Images}
\subsubsection{Observations and Reductions}

On February 25, 1991 we obtained $2 \times 300$ sec broadband I and $2
\times 600$ sec narrowband \ha exposures of NGC~3256, and E region
standard field \# 5~\citep{grah82}, with the 1 metre telescope at
Siding Spring Observatories (SSO).  The EEV CCD has $832 \times 1152$
pixels, pixel scale of 0.57 arcsec, readout noise 4 e$^{-}$ and gain 1
e$^-$ per adu.   The seeing was better than 2
arcsec. The central wavelength of the narrowband filter is 6594 \AA, its FWHM
52 \AA \ and its equivalent width 24 \AA.  At the redshift of NGC~3256,
the [NII]$\rm \lambda$6583\AA \ emission line lies in the wing of the
filter passband where its contribution to the image is negligible.

We first corrected the non-linearity of this CCD ($<$ 1.5\% at 18,000
adu to 6\% at the extrapolated value of 70,000 adu). Subtracting a fit
to the overscan region of every image reduced the low-level
time-dependent horizontal bias patterns caused by electronic pickup.
The dark current of this CCD is negligible so only bias images were
subtracted. The target images were flatfielded using dome and twilight
flatfields for each individual filter.  The processed images of the
galaxy were shifted to a common position and combined for each filter
with an algorithm that rejects cosmic rays. Image editing corrected
residual bad columns and pixels.

For the continuum images, we used the I-band, thus avoiding
contamination of the continuum images by \ha and [NII] emission
lines which are included in the R-band.  To scale the \ha and
I-band images, we used aperture photometry of the target galaxy
itself. Assuming that the outer regions (i.e. envelope) of the
galaxy disk do not contain a significant amount of \ha emission,
the I-band image was initially scaled such that the I-band
envelope matched the intensity of the envelope in the narrow-band
\ha image. We then experimented with scaling I to \ha continuum
within the inner 11 arcsec radius. We adopted the scaling which
avoided negative \ha flux in this H~II region zone when the
scaled I-band image was subtracted from the narrow-band image to
form a continuum-free \ha image.  This is likely to underestimate
the true \ha flux.

The slit spectra of the H~II regions have an equivalent width
in \ha $\rm \geq$ 80 $\rm AA$ confirming that they are bright. The
spectra also indicate that our underestimate in \ha photometry
is $\rm \leq$ 15 \%. 

\subsubsection{Flux Calibration}
\label{fluxcal}
The \ha flux density of our narrow band images was calibrated (in ergs
s$^{-1}$ cm$^{-2}$) using observations of 5 standard stars in the \#5
E-region,  with the zeropoints given by \nocite{bess90}(Bessell~1990,
\nocite{bess92}~1992).  The absolute flux density through the
narrowband  filter was determined by interpolating the broadband
zeropoints to the central wavelength  of the filter, and using the
equivalent width of the filter to calculate the integrated continuum
light for each standard star over the bandpass.

We used a circular aperture of 4 arcsec to measure the observed
sky-subtracted count rates for each star. These  were corrected for
atmospheric extinction using the mean airmass of the observations. The
mean extinction coefficient for the \ha filter, $k_{6596} = 0.085$,
was estimated by interpolation between the mean broadband coefficients
determined at SSO by Anja Schr\"{o}der (private communication). The
estimated error of this calibration is about 6\%.

\subsubsection{H~II Region Photometry}

Each irregularly-shaped giant H~II region complex in NGC~3256 was
outlined with a polygon. The sky-subtracted count rate, determined
from the annular sky aperture marked on Figure~\ref{photim}, was
corrected within each polygon for atmospheric extinction and for
Galactic absorption.  The Milky Way absorption ($\rm A_{6596}$ = 0.35
mag) was determined using the colour excess for
NGC~3256~\citep{burh82} and interpolating through average
interstellar extinction values $\rm A_{\lambda}$/E(B-V)~\citep{sm79}.
We did not attempt to correct for extinction within NGC~3256 itself.
This \ha flux was subsequently corrected for the transmission of the
filter at the wavelength (6625 \AA) associated with the redshift of
the galaxy (2820 \kms; see determination in \S~\ref{resultsvel}).  The
total \ha energy output, presented in Table~\ref{HIIreg}, was derived
using $\rm H_o = 100\ h \ km \ s^{-1} \ Mpc^{-1}$.

We did not attempt to divide up large complexes into apparent
subcomponents since our goal is to estimate whether the complexes are
comparable to 30 Doradus in their \ha flux.  The compromises in
polygon shape due to the proximity of other H~II regions provide the
largest uncertainty in the flux estimate ($\rm 1\ \sigma \sim$17\%), but
this is insignificant for our comparisons with the flux of 30 Doradus.

\begin{table}
\dummytable\label{HIIreg}
\end{table}

\subsection{\protect\ha Spectra} 
\subsubsection{Observations} 
In order to determine the velocities of the H~II regions, long slit
spectra were acquired at SSO using the Double Beam Spectrograph (DBS)
and Photon Counting Array (PCA) at the Nasmyth focus of the 2.3m
telescope on Mar. 8 \& 9, 1991.  The red 1200 l  $\rm mm^{-1}$ grating
gave a dispersion of  0.4 \AA\ ($\sim$ 19 \kms) per pixel, a FWHM
resolution  of about 1 \AA \ , and a total wavelength range of about
300 \AA.  The slit width was 1.8 arcsec and the seeing about 2 arcsec.
In the spatial direction, each pixel corresponds to 0.66 arcsec
resulting in a useful slit  length of  about 330 arcsec.

To place the slit precisely, we placed one end of the slit on one of 3
stars near the interacting system (labeled A, B, and C in
Fig.~\ref{havfld}). Using this star as a pivot point, the
spectrograph was rotated to various position angles on the sky.  For
each position angle a number of H~II regions lay along the slit and
hence produced a long-slit spectrum  which contained spectra for a
number of emission-line objects as well as the continuum of the
fiducial star (which provides a spatial reference point).

Exposures were typically 1000 sec. A neon lamp was observed before and
after every target spectrum.  This allowed us to correct for the small
flexure in the DBS as it rotated to maintain constant position angle
on the sky.  In order to ensure linearity of the detector in the
flatfield exposures, we took quartz lamp spectra at rates less than
0.1 Hz pixel$^{-1}$.

\subsubsection{\protect\ha Long-Slit Reductions}
\label{spec}
PCA data have no bias, and cosmic rays are rejected in hardware.
Therefore, before wavelength calibration, it was only necessary to 
flatfield the images as usual, using quartz lamp spectra: sky
flats provide the correction for vignetting along the slit. 

The wavelength calibrations, determining dispersion solutions and
rebinning to linear wavelengths, used standard procedures
~\citep{mvj92}.  We did not correct for distortions in the spatial
direction since these were insignificant at the accuracy required by
our study.  We subtracted the background only when night sky OH lines
lay very close to the redshifted H~II emission lines.

To measure the radial velocities, we summed over 3 rows ($\approx$ 2
arcec, to match the seeing) and determined the centre of the \ha
emission-line with a single-gaussian fitting routine. We then stepped
up one row, again summed 3 rows, and determined the central velocity of
the \ha emission at this new position.  This was repeated for all
sections in the 2-D spectrum containing emission.

The PCA has an intrinsic semi-periodic fringing pattern which changes
slowly with time.  Flatfielding removed most of this pattern. We
unsuccessfully attempted to remove the remaining fringing by 2D Fourier
filtering, and the residual coherent noise ultimately limited the
attainable S/N and dominated the error in our velocity measurements
which  increased with a decrease in emission intensity.  To estimate
this error empirically, we produced a calibration curve
using the sky lines in a set of night sky spectra.  A spline curve was
fit to a plot of (error in measured wavelength) versus (median
intensity of sky line).  This curve  indicated that spectral features
with peak intensities above 40 counts per wavelength pixel have an rms
error of 7 \kms \ (dominated by the wavelength calibration error).  The
error increases with decreasing counts to 18 \kms \ for peak
intensities  of about 10 counts per pixel.

\subsection{Velocity Field Analysis}
\label{optspec}

An optical velocity field image of H~II regions, Fig.~\ref{havfld},
was constructed by editing the \ha CCD image such that the
heliocentric velocity values were substituted at the image pixel
coordinates associated with the positions of the DBS slit.  Gaps in
the resultant DBS slit map, due to the difference in scale
(arcsec/pixel) between the DBS and CCD images, were assigned
interpolated velocity values. A typical positional uncertainty in the
velocity field image is about 1 CCD pixel (0.57 arcsec).  Due to the
long exposure times, the DBS spectra were more sensitive than the
imaging to diffuse emission.  This results in a velocity field data in
regions without H~II region peaks in the image.

The velocity field image was colour edited to heighten the contrast of
consecutive velocity value bins.  The range of colour bins is nearly
linear with each bin spanning 33.5 \kms.   The full colour range
was chosen to maintain the convention `red corresponds to redshift and
blue corresponds to blueshift'.\footnote {The saturation of the red
was set such that it was dull and the blue chosen was a `warm', bright
blue.  Hence the red colour, in terms of human visual perception,
appears to recede and the blue colour visually approaches, in the 
same sense as redshift and blueshift.}   To create the
intervening velocity range, the red and blue extrema were blended
together such that the neutral gray bin in Fig.~\ref{havfld} contains
the heliocentric systemic velocity value.

The data from the horizontal slit position (associated with fiducial
star A) is also presented in plot format (Fig.~\ref{harc}) in order to
provide an indication of the galaxy's velocity curve. Since the
inclination of this galaxy is close to face-on we have not attempted
to deproject the data.  Even when multiple gaussian features were
present in the 1-D spectra, they have been assigned an average
velocity; the errors given are flux dependent errors (see
\S~\ref{spec}) and do not reflect the range in velocity that these
features span.

Also in Table~\ref{HIIreg} we present the heliocentric velocities
corresponding to the point on the DBS slit which is closest to the
centre of each polygon used in the photometry (see Fig.~\ref{photim}).
The difference between a polygon's centre and the position of the peak
intensity of the H~II region can be up to a few arcsec.  Also, the
slit positions  generally did not pass directly through the centre of
the H~II region. A comparison of the two slits which pass near the
northern nucleus indicates that the velocities tabulated in
Table~\ref{HIIreg} may differ from the mean velocity of a given H~II
complex by up to 20 \kms.

\subsection{Relationship to Other Data}
These \ha images and spectra have in some respects been superseded 
by data acquired by \nocite{lip2000} L\'ipari \etal~(2000).  Their
\ha image from the ESO New Technology Telescope is also used by
\nocite{lira02} Lira \etal~(2002), who calibrated the flux (to
within a factor of 2) of the L\'ipari \etal image using the L\'ipari
\etal long slit spectroscopy of several H~II knotts.  However 
Lira \etal only report a total flux ($\rm \sim 2 \ \times \ 10^{-11}
\ erg \ s^{-1} \ cm^{-2}$ in an aperture with a 41 arcsec diameter).
L\'ipari \etal~(2000) measured the \ha flux of H~II knotts via
spectroscopy, effectively subdividing our selection of H~II regions
without measuring all the flux per region.  For example, they present
the flux of regions labeled R1, R2, and R4 but not of R8 and R9
although all of these would be in our regions B1 and B2; see
Table~\ref{correlate}. Although our measurements are therefore
difficult to compare, their R7 region's \ha flux differs from our
measurement of B6 by 8\%.

Our spectra are included not only for completeness sake but also because the
resolution is 34 \kms (with errors ranging from 7 to 18 \kms) compared
to the L\'ipari \etal~(2000) instrument resolution of 90 \kms (errors
ranging from 15 to 30 \kms).  Additionally L\'ipari \etal only present
velocity field maps with a width of 60 arcsec.  Researchers,
particularly those simulating the interactions between galaxies, may
find our 330 arcsec long slit provides useful information on almost
all radii associated with the ionized hydrogen emission component of
this galaxy and that our velocity field reveals interesting anomalies
beyond the inner region (e.g. redshifted emission in the tail to the
SW in Fig~\ref{havfld}.)

\section{Results}
\label{results}
\subsection{Velocity Curve and the Dynamical Mass}
\label{resultsvel}

Visual inspection of Fig.~\ref{havfld} shows largescale
rotation, in the same sense as the rotation observed in the neutral
hydrogen data (Paper~II). However the velocity
behaviour of the optical disk appears less disturbed by the merger
activity than do the H~I arms.

The peculiar morphology of NGC~3256, the uncertainty of its centre,
the position angle of the major axis, and its near face-on
inclination, make it difficult to extract an accurate deprojected
circular velocity curve.  Fig.~\ref{harc} shows a velocity curve from
a spectrum with the slit at position angle (P.A.) $90^{\circ}$ .
Examination of \ha features in Fig.~\ref{havfld} shows that the
turnover points in the velocity curve are representative of the
velocity range of the system.  The value at the midway point between
the turnover peaks is 2820 $\pm$ 11 \kms\ and the velocity difference,
$\rm \Delta V$, gives a rough estimate of the projected rotation
velocity; $\rm v_{circ} = \Delta V / 2 \approx 107 \ \pm 11 \ km \
s^{-1}$.

This agrees with the\nocite{fr78}Feast and Robertson (1978) estimate
of the systemic velocity and rotation amplitude (using P.A. =
$100^{\circ}$), \citet{lip2000} (using P.A.= $90^{\circ}$, however
only for an inner region 40 arcsec square), and with the H~I
position-velocity (PV) diagram shown in Paper~II.  (We note that the
distribution of H~I in the PV diagram means that the rotation
amplitude is significantly larger than the separation of the H~I horns
in the integrated H~I profile also shown in Paper~II.)

Assuming that the rotation curve is flat, we estimate that the mass
within radius r is ${\cal M}(r) = \rm 3.2 \times 10^4\, [v_{circ}(km \
s^{-1})]^{2}\, r(arcsec)\, h^{-1} \ \ {\cal M_{\odot} }$.  This is no
more than an order of magnitude estimate because the system is
unlikely to be in centrifugal equilibrium at this stage.  Dynamical
mass limits are given in Table~\ref{tabdynmass} for a few radii
associated with structural or kinematic features.

\begin{table}
\dummytable\label{tabdynmass}
\end{table}

\subsection{H~II Regions and Globular Clusters}
\label{phot}

Our goal in this section is to consider whether the H~II regions
observed in NGC~3256 could be GC progenitors. For an ionised hydrogen
region to be a GC progenitor, in the sense of \nocite{kc88}Kennicutt
and Chu~(1988), we expect it to have an \ha luminosity of at least
$\rm 1.5 \times 10^{40} \ ergs \ sec^{-1}$. This is comparable to the
total \ha luminosity of 30 Doradus and corresponds to an ionised
hydrogen mass of about $\rm 8.5 \times 10^5 \mo$~\citep{keni84}. We
refer to these giant H~II regions as \dghr.  Additionally, each \dghr\
should contain a rich star cluster; 30 Doradus harbours R136 which,
from HST images (\eg~\cite{mey93}), is known to have the the dense
core and extended halo structure typical of a young globular cluster
or YPC.  Since the timescale for the formation of a YPC is comparable
to the lifetime of a giant H~II region~\citep{kc88}, it is highly
likely that the cores of objects like 30 Doradus will be identified as
\y when the surrounding OB associations have faded.

HST observations of NGC~4038/9 (e.g. \nocite{ws95}Whitmore and
Schweizer~(1994)) show both spectacular populations of giant H~II
regions and hundreds of \y. Evidence of an association between \y and
\dghr\ is provided by a comparison of \y with multiwavelength emission
in NGC~4038/9 by \citet{zfw2001}.  They find that red clusters (age
$\rm \leq$ 5 Myr) are associated with  radio continuum, CO and
Infrared emission, that blue young clusters ($\rm 3 \leq \ age \ \leq
16\ Myr$) are correlated with \ha and Xray emission, and bright, older
clusters ($\rm 16 \leq \ age \ \leq 160\ Myr$) are correlated with
Xray emission.  They note that this is consistent with the cluster
heating the dusty environment it formed in, then ionizing the
surrounding gas, and finally expelling the surrounding material via
stellar winds and supernovae.

In NGC~3256 \citet{lira02} observed diffuse Xray emission, using
Chandra Xray Observatory, that spatially co-incides with diffuse \ha
emission and they claim that this is evidence that the Xray gas is
heated by a starburst (rather than an AGN).  Additionally most of
their discrete Xray sources are co-incident with H~II knotts; see
Table~\ref{correlate} for correlations with our observations of
diffuse emission. The potential sources generating the discrete Xray
emission include supernova remnants and Xray binaries.  This suggests
the sources are in clusters with massive stars, which may be initially
embedded in H~II regions in a manner consistent with the Zhang \etal
picture.
\begin{table}
\dummytable\label{correlate}
\end{table}

This picture is echoed by \citet{aah02} who specifically look for a
spatial correspondence between H~II regions and YPC in NGC~3256 using
NICMOS Pa$\rm _\alpha$ and Infrared images.  They use evolutionary
synthesis models to indicate that the $\sim$8\% coincidence rate,
compared to the number of H~II regions + star clusters, is due to
evolution effects and their detection thresholds.  That is, after 9
Myr the H-band luminosity, used to detect YPC, peaks while the \ha
luminosity drops below their detection limit.  Since the H-band
luminosity threshold allows detection of clusters from ages 0-100 Myr,
more YPC are detected than H~II regions.  Those H~II regions without
detected IR clusters probably harbour clusters which are obscured by
dust in the manner of Galactic H~II regions less than 5 Myr old.
\citet{aah02} believe that they are only detecting spatial
coincidences when the YPC have masses $\sim 10^6\mo$ and ages less
than 7 Myr.

At optical wavelengths \nocite{zaeks95}Zepf et al.~(1999) find about
1000 blue, compact clusters in the inner region of the NGC~3256
merger. Therefore, at an order of magnitude level, we can compare the
number of clusters like R136 to an estimate of the specific frequency
of \g required if NGC~3256 is a merger of 2 galaxies.

To determine the number of R136-like clusters we first
select only those Zepf \etal clusters with -0.55 $<$ (B-I)$<$ 0.5.
\nocite{bc93}Bruzual and Charlot~(1998) models generate unique ages
between 3 and 7 Myr for clusters in this color range.  As well,
models by \nocite{sk97} Schaerer and de Koter (1997 and papers
listed therein) suggest O and B stars provide ionizing radiation
for about 7 Myr, making these clusters comparable in age to the
observed H~II regions which might harbour them.

The observed absolute magnitudes ($\rm M_B$) of these blue \y are
calculated using $\rm H_o$ = 75 km $\rm s^{-1} \ Mpc^{-1}$.  The
change in the $\rm M_B$ of R136 (-12.6; \citet{kc88}) is also
calculated for the 3 to 7 Myr age range according to the
\nocite{bc93}Bruzual and Charlot~(1998) models.  Examining the
difference between the observed magnitude of a cluster and the
magnitude predicted for R136, at same age as the cluster, indicates
that 17 out of 196 of the bluest clusters have luminosities, and hence
masses, greater than R136.  The total mass in the 17 clusters is
equivalent to about 45 $\times$ the mass in R136 ($\cal M_{\rm
R136}$).  (The \citet{aah02} clusters that are coincident with H~II
regions probably have ages between 3-7 Myr and masses $\sim 10^6\mo$.
At least 18 coincidences occur in the 19.5 square arcsec field-of-view
of NIC2.  Scaling this area to match the 36 square arcsec area of the
PC chip predicts the existance of 60 IR clusters with R136-like
masses, which is comparable to our estimate of 45.) Extrapolating to
the $\sim$1000 \y detected by \citet{zaeks95} suggests that there are
roughly 100 \y with masses greater than R136 or a total of 225 $\cal
M_{\rm R136}$ clusters.

Thus we estimate the existence of 100 - 200 massive YPC's; 100 \y if
the masses are distributed amongst clusters in a manner similar to
this sample and approaching 200 \y if more of the clusters are about 1 $\cal
M_{\rm R136}$.  This is a conservative estimate since only
a limited field of view has been studied.  Additionally our
threshold of (B-I)$<$ 0.5 eliminates blue clusters reddened by
intervening dust and we note that all 17 clusters are situated in the
less dusty east section of the disk.

An alternative YPC estimate uses \ha photometry and also suggests that
more than 200 R136-like clusters could form over the lifetime of the
galaxy-galaxy interaction.  5 H~II complexes in Table~\ref{HIIreg}
have fluxes above $\rm 1.5 \times \ 10^{40} \ ergs \ sec^{-1}$, giving
a total of 85 \dghr\ not including those that might exist in the
obscured southern nucleus. (This is consistant with \citet{aah02} who
detect at least 50 \dghr\ in the FOV of NIC2, which includes only B1,
B5, part of B6 and B8. Our remaining H~II complexes contribute 67\% of
the \ha luminosity due to \dghr. This implies about 35 \dghr\ were not
imaged by NIC2.)  All the Zepf \etal clusters but one project onto the
sky in the vicinity of the 3 \dghr\ regions in the 7 kpc by 7 kpc
field of view that they analyzed.  The number of bright, blue
R136-like clusters associated with the \dghr\ suggests that 20-50\% of
these \dghr\ may currently cradle YPC's. Notably, \citet{aah02} claim
about 55\% of their high-luminosity H~II regions have an IR cluster
counterpart.

To estimate the total number of \y which might form from \dghr\ over
the era of the current interaction stage, we consider the rate of
\dghr\ formation in NGC~3256.  Using a lifetime of 7 Myr gives a
current rate of 12 Myr$^{-1}$ (compared to less than 0.2 Myr$^{-1}$
for a typical Sc galaxy with M$_{\rm b}$ = -19.5~\cite{kc88}).  This
enhanced rate could be maintained for longer than the age of the
nuclear starburst (between 10 and 27 Myr; \nocite{djw94}Doyon et
al.~(1994) and \nocite{riglutz96}Rigopoulou et al.~(1996)). However
the numerical merger simulations of Mihos \& Hernquist~(1994), in
which each parent galaxy has a bulge, suggest that this rate has
probably not been continuous.  The global star formation rate of each
parent galaxy increases by only a few percent until the galaxy cores
coalesce, and then the star formation rate increases dramatically.

Assuming that the cores are now about to merge, the time from
pericentre through to coalescence is about 50 model units,
corresponding to about 500 \ho Myr in NGC~3256 (see Paper~II).  The
peak of the star formation extends over about 4 model units, or about
40 \ho Myr in NGC~3256.  If we are now observing the peak star
formation rate, then this time interval generates at least 640 \dghr\ 
(for h = 0.75).  The formation rate for the previous 400 \ho \ Myr is
about 7\% of the peak rate value.  Hence another 250 H~II regions
probably formed between the time of pericentre separation and the
observed enhanced star formation epoch. If 20-50\% of the \dghr\
harbour R136-like clusters, then we might expect at least 200-400 \y
to form over the lifetime of the interaction.

Although there is some controversy about the relationship between the
YPC mass function and the GC mass function (see \citet{fritze01} and
papers therein), we believe that the R136-like clusters in the Zepf
\etal data will acquire GC colors and luminosities over several Gyr.
It is difficult to estimate how many of the R136-like clusters would
survive the evolution of the NGC~3256 \s \ over the next several Gyr.
However if the current power-law luminosity function of the YPC system
\citep{zaeks95} transforms to the log-normal distribution of an older
\s \ mainly via evaporation (due to internal relaxation) then the loss
of mass occurs mainly among clusters less massive than those in the
median mass bin~\citep{fallz01}. Since the mass of
R136 corresponds to the median bin values, we expect that these couple
of hundred \y may survive.

If typical ellipticals form via the merging of spirals, the
number of new massive clusters formed in the merger needs to be
at least comparable to the number of globular clusters originally
associated with the spirals~(\eg\ \citet{az93}).  Hence we now
estimate the original number of \g and compare it with the number
of \y in NGC~3256. To get an order of magnitude estimate of
the number of \g associated with NGC~3256's parent systems, we
arbitrarily assume that the parent galaxies have similar
bulge/disk ratios to M31, and compare the dynamical mass ($\rm
\sim 3 \times 10^{10} \ h^{-1} \ \mo$) within the radius of the
I-band envelope of NGC~3256 with the dynamical mass of M~31 ($
\rm 3 \times 10^{11}\ \mo$) within the 20 kpc radius of its
globular cluster system~(GCS)~\citep{huc93}.  Since M~31 has
about 400 confirmed \g ~\citep{fpcfp}, scaling by the ratio of
dynamical masses suggests that the original number of \g
belonging to the two-parent system was roughly 40-80.  The larger
value uses the \nocite{fr78}Feast and Robertson~(1978) inclination
and position angle (see \S3.1). As a check we use the specific
frequency per unit mass for disk galaxies (2.2 clusters $\times \ 
10^9 \ \mo^{\rm -1}$; \citet{za93}) which gives 66 \g.

In summary, since ellipticals have typically more than twice as many
\g per unit mass as spirals (e.g.~\citet{az93}), we need at least as
many \g to form in the merger as originally existed in the parent
system. If NGC~3256 is the result of the interaction of 2 similar disk
galaxies, as indicated by its 2 optical tails, then the number of \g
required would be less than 100. Extrapolating from the fraction of
the bluest \y observed in NGC~3256, roughly 100-200 \y are expected to
have masses equal to or greater than R136.  (Using \ha photometry and
numerical models suggests 200-400 R136-like clusters form.)  These
should fade to GC luminosities over several Gyr and many are expected
to survive destruction, since it is the lower mass clusters which
evaporate.  Hence the observed \y appear numerous enough to populate
the \s of the emerging galaxy such that its specific frequency will be
consistent with that observed in typical elliptical galaxies.

Additionally, the bluest, massive \y are located in the vicinity
of observed \dghr s, contributing to the notion that some
fraction of \dghr \ do cradle massive \y, as 30 Dor harbors
R136.  If interactions induce the formation of \dghr, the
observed luminosities indicate that almost 900 \dghr \ would form
in NGC~3256 throughout its merger epoch. In order for \dghr \ to
be considered GC progenitors, the specific frequency
argument only requires that about 10\%
of these harbor \y which survive several Gyr and have  
masses $\geq \ \cal M_{\rm R136}$.

\section{Conclusions}
\label{concl}

An approach, outlined in our introduction, that would be fruitful for
assessing whether giant H~II regions may be globular cluster birth
places, uses hydrodynamic modeling in order to see if the kinematics
of the H~II regions at various interaction and merger stages is
consistent with the final spatial distribution of \g in elliptical
(i.e. merger remnant) galaxies.  This approach should be applied to a
sample of galaxies spanning the interaction stages that lead to a
merger remnant. It requires a comparison of observed H~II region
positions and velocities throughout the merger sequence with the
statistical position and velocity distributions of condensations of
stellar and gas particles seen in simulations which generate
elliptical-like merger remnants (e.g.  \nocite{mh96} Mihos \&
Hernquist~(1996)).  At every interaction stage, the behaviour of the
observed progenitor candidate should be similar to that of the
condensations in the model. If this picture is correct both the
observed and numerical sequence should produce a GCS that is
consistent with GCS's observed around elliptical galaxies.

Another approach compares the observed specific frequency of \g in
typical ellipticals with an estimate of the number of \g formed as two
similar disk galaxies interact and coalesce to form an elliptical
merger remnant.  Since ellipticals have typically twice as many \g per
unit mass as spirals, at least as many \g need to be created in the
merger process as originally existed in the sum of the 2 parent
galaxies \citep{az93}.  The argument is that tidal
disturbances provide a mechanism for enhancing star formation and
hence generate giant H~II regions in the disk of the emerging
elliptical.  On theoretical grounds interactions are expected to lead
to enhanced H~II region formation~(e.g. \citet{lar87}), and
observations demonstrate that giant H~II region populations are
enhanced at various stages of the interaction-through-to-merger
sequence (e.g.  NGC~4038/9, NGC~5426).  These ionised complexes in
turn ``cradle'' YPC's.  (Support that \y are associated with H~II
regions is provided by their correlation in
NGC~4038/39~\citep{zfw2001}.)  Massive \y are expected to survive
evaporation from the GC system of the emerging elliptical and we
believe that these \y will fade over several Gyr such that they would be
recognized as GC's.The optical data on NGC~3256 presented in this
paper allows us to explore this approach.

To get an order of magnitude estimate of the number of \g associated
with NGC~3256's parent systems, we arbitrarily assumed that the parent
galaxies have similar bulge/disk ratios to M31. Using M31's number of
GC per unit mass suggests the original number of \g belonging to the
two-parent system was roughly 40-80. Thus over NGC~3256's 
merger period about 100 \y must form and survive, if the above scenario
is correct.  

Although NGC~3256 has relatively few (7) H~II complexes compared to
some other interacting systems, 5 of these regions are comparable in flux
to about 85 30-Dor giant H~II regions (\dghr; each with an \ha flux of
1.5 $\rm \times 10^{40} \ erg \ sec^{-1}$).  Given
that these \dghr\ formed within the previous 7 Myr, this merging
system currently has a giant H~II region formation rate 60 times that
observed in an Sc galaxy with M${\rm _B} = -19.5$ \citep{kc88}. The 30
Dor nebula contains the YPC R136 within NGC~2070 \citep{mey93} and the
formation timescale for \y is comparable to the lifetime of a giant
H~II region \citep{kc88}.  As well, clusters as massive as R136 are
expected to survive evaporation due to internal relaxation
(S.~M.~Fall, private communication) and hence become part of the
remnant elliptical's GCS.  Therefore we used this flux as a minimum
criterion for identifying which H~II regions are potential birth
places of \y~\citep{kc88}.
 
An order of magnitude estimate of the total number of \y created
throughout the merger sequence experienced by NGC~3256's parent
galaxies was determined 2 ways.  For one estimate we determined that
about 1/10 of the $\sim$ 200 bluest clusters observed in the inner
region of NGC~3256 by \citet{zaeks95} have luminosities, and hence
masses, {\em greater than} R136. Extrapolating to the $\sim 1000$
detected clusters implies 100-200 \y exist. The alternative estimate
used \ha photometry and the variation in star formation rate over the
interaction-through-merger timescale, calculated using existing
numerical simulations\citep{mh94}. These indicated about 900 \dghr\
formed since the last pericentre approach of the parent galaxies,
yielding 200-400 \y if 20-50\% of \dghr\ have R136-like clusters.  We
expect clusters this massive to survive destruction.

If NGC 3256 is typical of the merging systems that form globular
clusters, and if about 10\% of the \dghr\  harbour clusters
that survive several Gyr, this would sufficient to explain the
high specific frequencies of clusters in ellipticals.

\bigskip

We are grateful to Mike Dopita for suggestions on observing strategy,
to Peter Quinn for perspective on interactions and mergers, to
Peter Wood for the use of his CCD image linearizing script, and to
the technical staff at Mount Stromlo and Siding Spring Observatories. 
J.~E. acknowledges the support of an Australian National University
Postgraduate Scholarship.

\newpage

\figcaption[EngFree.fig1.ps]{Photometry of H~II regions. The
underlying \protect\ha image is continuum subtracted.  The annulus
marks the region from which the mode sky value per pixel was
determined. The polygons mark the areas measured. Those designated B1
through B7 are treated as H~II regions and are described in Table
~\protect\ref{HIIreg}. Comparisons with positions of \ha emission
knotts and X-ray sources in other projects are in Table
~\protect\ref{correlate}.  The area of B8 lies close to an
optically-obscured radio source. The inner radius of the annulus
is 43 arcsec. \label{photim}}

\figcaption[EngFree.fig2.ps]{The velocity field of the ionised gas in
NGC~3256. The underlying image is the ionised hydrogen data acquired
using the SSO 1 metre telescope.  The spectroscopic data was acquired
using the SSO 2.3 metre telescope, the Double Beam Spectrograph (DBS),
and the Photon Counting Arrays (PCA). Each `line' consists of the
velocity values at a slit position. Red indicates receding velocities;
blue indicates approaching velocities. The construction of this map is
detailed in \S~\protect\ref{optspec}.  The stars used as fiducial
pivot points for the DBS slit positions are labeled.\label{havfld}}

\figcaption[EngFree.fig3.ps]{Ionised hydrogen velocity curve.  The DBS
slit at position angle $90^{\circ}$ and anchored at star A; see
Figure. 2.  The declination of this observation is north of northern
nucleus.  The rms error in velocity, determined using the calibration
curve described in \S~\protect\ref{spec}, is 7 \kms \ between 10h 27m
49.7s and 10h 27m 53s.  Outside this R.A. range the counts decrease
and the error increases to 18 \kms.
\label{harc}}

\end{document}